\documentclass[11pt,a4paper]{article}
\usepackage{jheppub}

\usepackage{bbold}

\usepackage[english]{babel}

\newcommand{\pmat}{\begin{pmatrix}}
\newcommand{\fpmat}{\end{pmatrix}}
\newcommand{\eq}{\begin{equation}}
\newcommand{\feq}{\end{equation}}
\newcommand{\cas}{\begin{cases}}
\newcommand{\fcas}{\end{cases}}

\newcommand{\eqarray}{\begin{eqnarray}}
\newcommand{\feqarray}{\end{eqnarray}}

\newcommand{\be}{\beta}

\def\be{\begin{equation}}
\def\ee{\end{equation}}
\def\bea{\begin{eqnarray}}
\def\eea{\end{eqnarray}}

\title{String Memories ... openly retold}
\author{A. Aldi,}
\author{ M. Bianchi,}
\author{M. Firrotta}
\affiliation[a]{Dipartimento di Fisica, UniversitÃ  di Roma Tor Vergata\\
Via della Ricerca Scientifica 1, 00133, Roma, Italy}
\affiliation[b]{INFN sezione di Roma Tor Vergata \\
Via della Ricerca Scientifica 1, 00133 Roma, Italy}

 \emailAdd{alice.aldi@roma2.infn.it}
 
 \emailAdd{massimo.bianchi@roma2.infn.it}
 
 \emailAdd{mfirrotta@roma2.infn.it}

\abstract{We identify string corrections to the EM memory effect. Though largely negligible in the low-energy limit, the effect become relevant in high-energy collisions and in extreme events. We illustrate our findings in a simple unoriented bosonic string model. Thanks to the coherent effect of the infinite tower of open string resonances, the corrections are non-perturbative in $\alpha'$, modulated in retarded time and slowly decaying even at large distances from the source.
Remarkably compact expressions obtain for special choices of the kinematics in tree-level 4-point amplitudes. We discuss further corrections occurring at higher-points and the exponential damping resulting from broadening and shifting of the massive poles due to loops.  Finally we estimate the range of the parameters and masses for detectability in semi-realistic (Type I) contexts and propose a rationale for this string memory effect.}

\makeatletter
\gdef\@fpheader{}
\makeatother

\begin{document}
\maketitle

\section*{\emph{Introduction}}

Thanks to the universal behaviour of soft photons and gravitons \cite{OldSoft}, a memory effect is expected to take place both in electro-dynamics \cite{OldEMMemo} and in geometro-dynamics \cite{OldGravMemo}. 

In a series of papers \cite{StromTri} a triangle of equivalences Soft Behaviour \cite{OldSoft, NewSoft} - BMS Symmetry \cite{OldBMS, Barnich, NewBMS} - Memory Effect \cite{NewMemo} has been put forward and proposals for experimental tests have been suggested \cite{GravMemoExp}.  
While gravitational memory implies a distortion of the detector after a GW has passed through \cite{OldEMMemo}, EM memory corresponds to a residual velocity (called a kick) for the charged particles of the detector \cite{OldEMMemo}. 

Since String Theory is a consistent quantum theory of gravity and electro-magnetism, one expects a similar story to be told. In fact, the leading soft behaviour of string scattering amplitudes is the same as in field theory\cite{StringSoft}. Since the soft behavior determines the profile of electro-magnetic ({{EM}}) and gravitational waves (GW) at large distances from the source, one would naively expect no string corrections to the memory effect\footnote{The term `string memory effect' was coined in \cite{Afshar1811.07368} in relation to large gauge transformations of $B_{MN}$.}. Indeed standard low-energy expansions of string scattering amplitudes would produce corrections in $\alpha'$ that are highly suppressed at large distances and would be totally negligible. However, taking into account the infinite tower of string resonances changes the story completely. 

The coherent effect of string resonances, that are crucial for the finiteness and consistency of the theory, play a key role in the non-trivial corrections to the GW profile produced in a `stringy' BH merger in the heterotic string \cite{ABFMhet}. Here we argue that a similar phenomenon takes place for {{EM}} waves in the Veneziano model a.k.a. open bosonic strings. Actually we consider a variant that requires internal dimensions and Wilson lines \cite{MBASsys, MBASWL, BPStor} or in modern language D-branes and $\Omega$-planes \cite{AS, GPAS, Polch}.

We focus on the simplest non-trivial scattering amplitudes on the disk with insertion of a single photon\footnote{Non-linear aspects of stringy EM, encoded in the DBI action for D-branes, only play a role in multi-photon emission. We thank Cobi Sonnenschein for raising this issue.}, which expose a pole in the frequency at $\omega =0$, reproducing the {{EM}} memory effect, and a series of simple poles on the real axis, related to the open string resonances. 

Integration over $\omega$ produces string corrections to the {{EM}} wave profile even at large distances from the collision $R>>L$. The $1/R$ terms form a series in ${{\zeta}}=\exp(-i{{u}}/{{\ell}})$, where $u=t-r$ is the retarded time and $\ell= 2\alpha' n{{\cdot}}p$, with $n= (1, \vec{x}/R)$ and $p$ one of the momenta of the charged particles involved in the process. The series can be summed for special kinematics {\it i.e.~} for `rational' values of the ratios $\lambda_{i,j} = n{{\cdot}}p_i/n{{\cdot}}p_j$.    

Clearly for $u{>>}{{\ell}}$ {\it i.e.} at low energy the effect averages to zero due to the strong oscillations but we expect a detectable signal for $u\approx {{\ell}}$, {\it i.e.} in high-energy processes with (many) heavy charged particles, such as collisions of cosmic strings or BH mergers. Since the effect is general and takes place in more realistic contexts, such as chiral Type I models \cite{ABPSS}, we will estimate the order of magnitudes for the phenomenon to be observed in extreme processes where stringy structures may play a role. We will also propose a rationale for the origin of this string memory effect. 

Let us start with the general line of arguments leading from the classical EM memory to its stringy version and then illustrate the phenomenon in the simple context of unoriented open bosonic strings, a close relative to Veneziano model\footnote{More details and more elaborate examples will be given in a companion paper \cite{Compa}.}.

\section{ \emph {EM memory and string corrections}}

In classical electro-dynamics, the {{EM}} field produced by a source ${{J}}_{\mu}$ is given by retarded potential $A^{{\rm ret}}_{\mu}$. 
In Fourier space, at large distances from the source $R=|\vec{x}|>\!\!> |\vec{x}'|\approx L$, setting $\vec{n}= \vec{x}/R$, one finds\footnote{We denote by $\widetilde{G}(\omega,\vec{x})$ the Fourier transform w.r.t. $t$ and by  $\widehat{G}(\omega,\vec{k})$ the further transform w.r.t. $\vec{x}$.}
\be
\label{tildeAfromJ}
\widetilde{A}^{\mu}(\omega, \vec{x})=\int d^3x' {e^{{{i}} \omega|\vec{x}{-}\vec{x}'|} \over 4\pi |\vec{x}{-}\vec{x}'|} 
    \widetilde{{J}}^{\mu}(\omega,\vec{x}') 
\approx { e^{{{i}} \omega R} \over  4\pi R} 
\widehat{{J}}^{\mu}(\omega,\vec{k}=\omega\vec{n})\,.
\ee


In QED, the leading behavior (as $k\rightarrow 0$) of an amplitude with a soft photon and $n$ hard particles with charge $Q_a$ is dictated by \cite{OldSoft} 
\be 
{\cal A}^{QED}_{n{+}1}(a,k;p_j) = {g}
\sum_{j=1}^n {Q_j a{\cdot}p_j  \over k{\cdot}p_j}{\cal A}^{QED}_{n}(p_j)+ ...
\ee
where $g$ is the charge quantum. 
Stripping off the photon polarisation $a^\mu(k)$, the amplitude becomes a source for the `classical' {{EM}} potential 
that at large distances assumes the form  
\be 
\widetilde{{A}}^{\mu}(\omega, \vec{x}) = {g}  {e^{{{i}} \omega R} \over \omega R}\sum_j {Q_j p_j^{\mu}\over np_j}  \,,
\ee
with $n^\mu{=}k^\mu/\omega = (1, \vec{x}/R)$. Integrating over $\omega$, the pole at $\omega=0$ produces a constant shift of ${{A}}^{>}_\mu(t,\vec{x})$ at late retarded time $u=t-R$ w.r.t. ${{A}}^{<}_\mu(t,\vec{x})$, which is known as `electro-magnetic memory'. 


In String Theory (ST), whenever a massless abelian vector boson is present in  the spectrum as in Veneziano model, the (transverse) `current' that sources the EM potential ${{A}}_{\mu}$ is given by  
\begin{equation}
\widehat{J}_\mu^{ST}(k;p_j) = {\delta {\cal A}^{ST}_{n{+}1}(a,k;p_j)\over \delta{a}^\mu(k)} \, .
 \end{equation} 

In the low-energy limit, Veneziano amplitude or its generalisations can be expanded in powers of $\alpha' k{{\cdot}}p_j$. Fourier-transforming in $\omega$ back to $t$ would produce string corrections to the retarded potential decaying faster than $1/R$, that would be totally negligible at large distances. 

On the other hand, for $\alpha' k{{\cdot}}p_j\approx 1$, including the contribution of the infinite tower of string resonances turns out to produce a coherent effect that is non-perturbative in $\alpha'$ and corrects even the leading $1/R$ terms.  Indeed, starting from  
\be
\widetilde{{A}}^{\mu}(\omega, \vec{x})= \int d^3x' {e^{{{i}} \omega|\vec{x}{-}\vec{x}'|} \over 4\pi |\vec{x}{-}\vec{x}'|} 
    \widetilde{J}^{\mu}(\omega,\vec{x}'; p_j) 
\approx { e^{{{i}} \omega R} \over  4\pi R} 
{\delta {\cal A}^{ST}_{n{+}1}(a,k;p_j)\over \delta{a}_\mu(k)}\Big\vert_{k^\mu=\omega(1,\vec{n})}
\ee
and integrating over $\omega$ yields contributions from the tower of poles lying on the real axis (at tree level). These form various series in the variables ${{\zeta}}_j=\exp(-i{{u}}/2\alpha' n{{\cdot}}p_j)$, that may give rise to detectable signals 
\be
{{A}}^{ST}_{\mu}(t, \vec{x}) = {{A}}^{QED}_{\mu}(t, \vec{x}) + \Delta_s{{A}}_{\mu}(t, \vec{x}) \, .
\ee
Contrary to the `standard' EM memory, which is a DC effect relating the behavior at $u>0$ to the one at $u<0$, the `EM string memory' is modulated {\it i.e.} depends on $u$. Quite remarkably $\Delta_s{{A}}_{\mu} = \theta(u) \Delta_s{{A}}^{>}_{\mu} + \theta(-u) \Delta_s{{A}}^{<}_{\mu}$ inherits peculiar duality properties from the parent string amplitudes {\it i.e.} $\Delta_s{{A}}^{>}_{\mu} + \Delta_s{{A}}^{<}_{\mu}=0$. 

In theories with open unoriented bosonic strings, massless vector bosons are ubiquitous. At tree level (disk) `color-ordered' amplitudes\footnote{We only consider open-string insertions on the boundary. Amplitudes with closed-string insertions on the bulk have been recently studied in \cite{DiskAmp}.} are cyclic and expose simple poles in channels corresponding to sums of consecutive momenta.
In particular soft poles in $kp$ arise as usual when a photon is inserted on a charged leg. This is accompanied by `massive' poles $1/2\alpha' kp{+}n$ that are responsible for the string corrections to the {{EM}} memory. Complete amplitudes require summing over non-cyclic orderings and expose all the expected  soft poles as well as massive ones. Multi-particle channels involving a photon, that give rise to sub-leading corrections to the soft behavior in field theory, produce new towers of massive poles and further string corrections to the {{EM}} memory.   

Actually, massive string resonances are unstable and acquire finite width and mass-shifts due to loop effects. As a consequence an exponential damping of the string memory will result \cite{ABFMhet}. Still, assuming $g_s{<<}1$, this may be a small effect that can be taken into account in a detailed analysis of the signal, very much as QNM's with ${\rm Im}\omega \neq 0$ in the Ring-down phase of BH mergers.

\subsection{ \emph {Setup}}
Let us identify a convenient setup with a massless photon and two charged (massive) scalars that allow to illustrate the EM string memory. To this end we compute a 4-point amplitude that admits a closed-form expression for special kinematics. We then extend our analysis to higher points, where new structures appear and briefly address higher loops that mark the onset log corrections. 

In the presence of an $\Omega$25-plane, dilaton tadpole cancellation  requires the introduction of $2^{13}$ D25-branes, resulting in the gauge group $SO(8192)$ \cite{MBAS8192, Alt8192}. Neglecting closed strings and the coupling to gravity, one can work in a local setting with D3-branes and $\Omega$3-planes well separated from the remaining branes. By T-duality this is equivalent to a configuration of D25-branes with Wilson lines \cite{MBASsys, MBASWL, BPStor}. 

\begin{figure}[h!]
\centering
\includegraphics[scale=0.3]{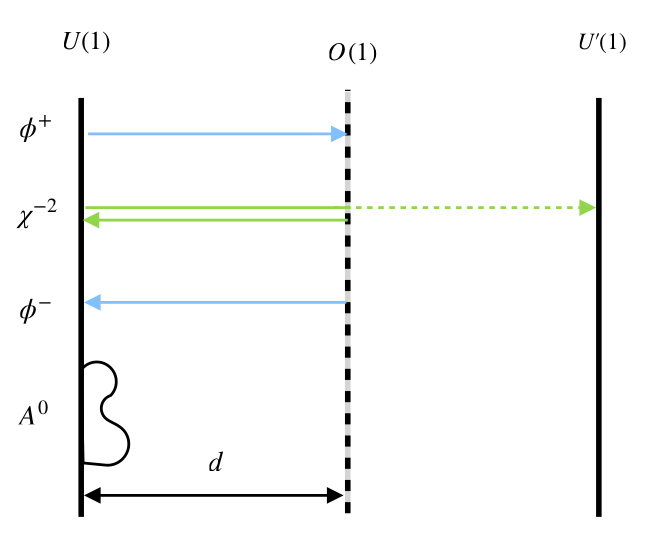}
\caption{Representative picture of the $\Omega3/D3$ setup.}
\label{F1}
\end{figure}

More specifically we consider a 4-dimensional configuration (Fig.~\ref{F1}) with one D3-brane on top of an $\Omega$3-plane, giving rise to an $O(1)$ gauge group (no vector bosons), and one D3-brane (together with its image) parallel to the $\Omega$3-plane and separated from it by a distance $d$ in one of the 22 `internal' directions giving rise to  a $U(1)$ gauge group with a  massless photon like in Maxwell theory\footnote{Actually, after coupling to closed strings, the photon mixes at the disk level and will be eaten by the $B_{MN}$ or else will eat a `dilaton'. As already said, we work at open-string tree-level and safely neglect this complication.}. 

In addition to the massless photon, the low-mass spectrum contains neutral and charged tachyons. Discarding the neutral ones (a `symmetric' singlet of $O(1)$ and the neutral singlet of $U(1)$), we have a singly charged scalar  $\phi^+$ and  its conjugate $\phi^-$, stretching from the $U(1)$ D3-brane to the $O(1)$ D3-brane and {\it vice versa}, with mass $\alpha' M_{\pm 1}^2 = -1 + {{\delta}}^2$ (where ${{\delta}}^2 = d^2/\alpha'$)  and a doubly charged  scalar $\chi^{+2}$ and  its conjugate $\chi^{-2}$, stretching from the $U(1)$ D3-brane to its image and {\it vice versa}, with mass $\alpha' M_{\pm 2}^2 = -1 + 4{{\delta}}^2$. For $\delta>1$ the `tachyons' are massive. We safely assume this to be the case and largely neglect the extra dimensions henceforth.

The vertex operators for $\phi^{+}$ and $\chi^{-2}$ take the standard `tachyonic'  form $\mathcal{V}_\phi= \sqrt{2\alpha'}e^{iK{\cdot} X}$ with $K=(p^\mu; \pm\vec{d})$ for $\phi^{\pm}$ and $K=(p^\mu; \pm 2\vec{d})$ for $\chi^{\pm 2}$. 
For the $U(1)$ gauge boson $A_\mu$ one has
$
\mathcal{V}_{A} = a{\cdot} i{\partial} X e^{ik{\cdot} X}
$  
with no `internal' components of the momentum. 

At tree level (disk) non-vanishing 3-point amplitudes are:
${\cal A}_{\phi^{\pm}\phi^{\pm}\chi^{\mp2}}=g_{op}/\sqrt{2\alpha'}$, where $g_{op}{=}\sqrt{g_s}$ is the open string coupling, and the `minimal' couplings of $\phi^{\pm}$ and $\chi^{\mp2}$ to the photon ${\cal A}_{A\phi_{_Q}\phi_{_{-Q}}} =  Q g_{op} a{{\cdot}}(p_1{-}p_2)$.

\subsection{\emph {4-pt amplitude}}

As a 4-point amplitude with a single photon insertion, one can consider 
\begin{figure}
\centering
\includegraphics[scale=0.4]{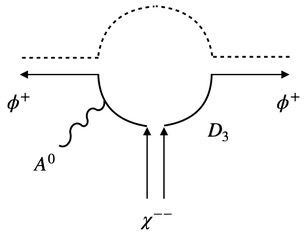}
\includegraphics[scale=0.4]{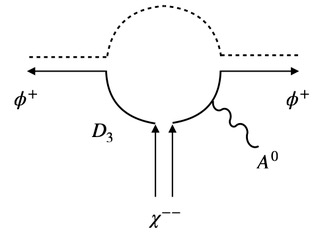}
\caption{Amplitude with one $U(1)$ photon and 3 `tachyons'.}
\label{Amp3+1}
\end{figure}
\be
{\cal A}_{3{+}1}= {{g}}^{4}_{op}{\cal C}_{D2}\prod_{j=0}^{3} \int\frac{
dz_{j}}{V_{CKG}}\left<\mathcal{V}_{A}(k,z_{0})\mathcal{V}_{\phi^{+}}(K_{1},z_{1})\mathcal{V}_{\phi^{+}}(K_{2},z_{2})\mathcal{V}_{\chi^{-2}}(K_{3},z_{3})\right>
\label{amp1}
\ee
where ${\cal C}_{D_{2}}=(g_{op}2\alpha')^{-2}$. Including the two contributions ${\cal A}_{A\phi\phi\chi}$ and ${\cal A}_{\phi \phi A\chi}$ in Fig.~\ref{Amp3+1} yields
\be
{\cal A}_{3{+}1}={g^{2}_{op}\over\sqrt{2\alpha'}} \left\{\left( \frac{a{\cdot} p_{1}}{k{\cdot} p_{1}} - \frac{a{\cdot} p_{3}}{k{\cdot} p_{3}}\right)\frac{\Gamma(2\alpha' k{\cdot} p_{1}+1)\Gamma(2\alpha' k{\cdot} p_{3}+1)}{\Gamma(1-2\alpha' k{\cdot} p_{2})} + (1\leftrightarrow 2)\right\}\\
\ee
Setting $k=\omega n=\omega(1, \vec{n})$ and defining the `scattering lengths' \cite{ABFMhet}
\be
{{\ell}}_{a}=2\alpha' np_{a}\qquad {\rm :} \qquad \sum_{a=1}^{3} {{\ell}}_{a} = 0
\label{elle} \: , 
\ee
thanks to $\sum_{a=1}^{3} k{\cdot} p_{a} = -k^{2} = 0$, the amplitude can be written as
\be
{\cal A}_{3{+}1} = g^{2}_{op}\sqrt{2 \alpha'} \left\{
\left( \frac{a{\cdot} p_{1}}{{{\ell}}_{1}} - \frac{a{\cdot} p_{3}}{{{\ell}}_{3}}\right) {\cal H}_{13}(\omega) + (1\leftrightarrow 2) \right\}
\label{amp5}
\ee
where the function (a generalisation of Veneziano amplitude)
\be
\mathcal{H}_{13}(\omega) = {1\over \omega} \frac{\Gamma(1{+}\omega {{\ell}}_{1})\Gamma(1{+}\omega {{\ell}}_{3})}{\Gamma(1{-}\omega {{\ell}}_{2})}\label{funomega}
\ee
has simple poles at $\omega=0$ as well as at 
$\omega {{\ell}}_{1} = {-}n_{1}{-}1$ and $\omega {{\ell}}_{3} ={-}n_{3}{-}1$ and admits an expansion  \`a la Mittag-Leffler (ML) of the form
\be
\label{MLexp}
\begin{aligned}
\mathcal{H}_{13}
=& \frac{1}{\omega} + \sum_{n_{1}=1}^{\infty}\frac{(-1)^{n_{1}}{{\ell}}_{1}}{n_{1}!( \omega {{\ell}}_{1} +n_{1})}\frac{\Gamma(1-n_{3}\lambda_{3,1})}{\Gamma(1+n_{1}\lambda_{2,1})} + \sum_{n_{3}=1}^{\infty}\frac{(-1)^{n_{3}}{{\ell}}_{3}}{n_{3}!( \omega {{\ell}}_{3} +n_{3})}\frac{\Gamma(1-n_{3}\lambda_{2,3})}{\Gamma(1+n_{3}\lambda_{1,3})}
\end{aligned}
\ee
where the ratios \be
\lambda_{{b},{a}}= \frac{{{\ell}}_{b}}{{{\ell}}_{a}} = {n{{\cdot}}p_b\over n{{\cdot}}p_a} = 
{k{{\cdot}}p_b\over k{{\cdot}}p_a}\,.
\ee
satisfy $\lambda_{3,1}{+}\lambda_{2,1}{=}{-}1$ and cyclic. The same applies to $\mathcal{H}_{23}(\omega)$ after $1\leftrightarrow 2$ exchange.

For the EM wave profile at large distances one has 
\be 
\widetilde{{A}}^{\mu}(\omega, \vec{x})= g_{op}\, \widehat{{\cal A}}_{3}(p_j) 
 {e^{{{i}} \omega R}\over 4\pi R} \sum_j {\,Q_j p_j^{\mu} \over  n{\cdot}p_j} {\cal F}_j (\omega,\vec{n}; p_j) \,,
\ee
where ${\cal F}_1= {\cal H}_{13}$, ${\cal F}_2= {\cal H}_{23}$ and ${\cal F}_3= {1\over 2}({\cal H}_{13}+{\cal H}_{23})$ and $\widehat{{\cal A}}_{3} = 2g_{op}/\sqrt{2 \alpha'}$ is an overall, non-zero factor, that can be absorbed into the largely unknown distance $R$ from the source.
Anti-Fourier transforming, one finds
\be 
\label{after}
{{A}}^{\mu}(t, \vec{x})=
{{g_{op}}\over 4\pi R} \sum_j {Q_j p_j^{\mu} \over {n{\cdot}{p}_j}} \int_{-\infty}^{+\infty} {d\omega \over 2\pi } e^{- {{i}} \omega u} {\cal F}_j(\omega,{{\ell}}_j) \, .
\ee
The pole at $\omega{=}0$ reproduces the {{EM}} memory DC effect. In addition to this, one finds genuine (open) string corrections $\Delta_s A^{\mu}(t, \vec{x})$ to the retarded potential. Adopting some reasonable prescription to deform the integration path, {\it i.e.} $kp_a{\rightarrow} kp_a{-}{{i}} \epsilon$, one can perform the integral and get series in ${{\zeta}}_j = e^{iu/\ell_j}$. Note that the effect is non-perturbative in $\alpha'$ as $\ell_j=2\alpha' n p_j$.
Assuming $\phi^+(p_1)$ and $\phi^+(p_2)$ to be incoming ($p^\mu{=}{-}p^\mu_{\rm phys}$) and $\chi^{-2}(p_3)$ and $a(k)$ to be outgoing ($p^\mu{=}{+} p^\mu_{\rm phys}$), in the physical kinematic region one has ${{\ell}}_{1,2}>0$ and ${{\ell}}_3<0$. 
For $u=t-R >0$ one finds
\be 
\begin{split}
\Delta^{(>)}_s{{A}}^{\mu}(t, \vec{x}) = -{{g} \over 4\pi R}&\Bigg\{ { p_1^{\mu} \over n{{\cdot}}p_1} \sum_{n_3=1}^\infty 
{(-)^{n_3} \over n_3!} {\Gamma(1{-}n_3\lambda_{1,3})
\over  
\Gamma(1{+}n_3\lambda_{2,3})} \,e^{{{i}} n_3 u/{\ell}_3} \\
&+ { p_2^{\mu} \over n{{\cdot}}p_2} \sum_{n_3=1}^\infty 
{(-)^{n_3} \over n_3!} {\Gamma(1{-}n_3\lambda_{2,3})
\over  
\Gamma(1{+}n_3\lambda_{1,3})} \,e^{{{i}} n_3 u/{\ell}_3}\\
&- { p_3^{\mu} \over n{{\cdot}}p_3} \sum_{n_3=1}^\infty 
{(-)^{n_3} \over n_3!} \left({\Gamma(1{-}n_3\lambda_{1,3})
\over  
\Gamma(1{+}n_3\lambda_{2,3})} {+}{\Gamma(1{-}n_3\lambda_{2,3})
\over  
\Gamma(1{+}n_3\lambda_{1,3})}  \right)\,e^{{{i}} n_3 u/{\ell}_3} \Bigg\}
\end{split}
\label{A>}
\ee
while for $u=t-R <0$ one finds
\be
\begin{split}
\Delta^{(<)}_s{{A}}^{\mu}(t, \vec{x}) ={{g}\over 4\pi R} & \Bigg\{ { p_1^{\mu} \over n{{\cdot}}p_1}
 \sum_{n_1=1}^\infty 
{(-)^{n_1} \over n_1!} {\Gamma(1{-}n_1\lambda_{3,1})
\over \Gamma(1{+}n_1\lambda_{2,1})} e^{{{i}} n_1 u/{{{\ell}}_1}} \\ 
&+{p_2^{\mu} \over n{{\cdot}}p_2}
 \sum_{n_2=1}^\infty 
{(-)^{n_2} \over n_2!} {\Gamma(1{-}n_2\lambda_{3,2})
\over \Gamma(1{+}n_2\lambda_{1,2})} e^{{{i}} n_2 u/{{{\ell}}_2}} \\
&\hspace{-2cm}-{p_3^{\mu} \over n{{\cdot}}p_3} \left(
 \sum_{n_1=1}^\infty 
{(-)^{n_1} \over n_1!} {\Gamma(1{-}n_1\lambda_{3,1})
\over \Gamma(1{+}n_1\lambda_{2,1})} e^{{{i}} n_1 u/{{{\ell}}_1}} + \sum_{n_2=1}^\infty 
{(-)^{n_2} \over n_2!} {\Gamma(1{-}n_2\lambda_{3,2})
\over \Gamma(1{+}n_2\lambda_{1,2})} e^{{{i}} n_2 u/{{{\ell}}_2}}  \right)\Bigg\} 
\end{split}
\label{A<}
\ee
Recall that the dependence on the position is coded in $n{{\cdot}}p_j{=}{-} E_j(1{-}\vec{n}\vec{v}_j)$ with $\vec{n}{=}\vec{x}/R$.  
The series have finite radii of convergence that may exclude the physical domain $|{{\zeta}}_{{j}}|{=}1$. Yet they can be summed explicitly for special choices of the kinematics. 

\begin{figure}[h!]
\centering
\includegraphics[scale=0.4]{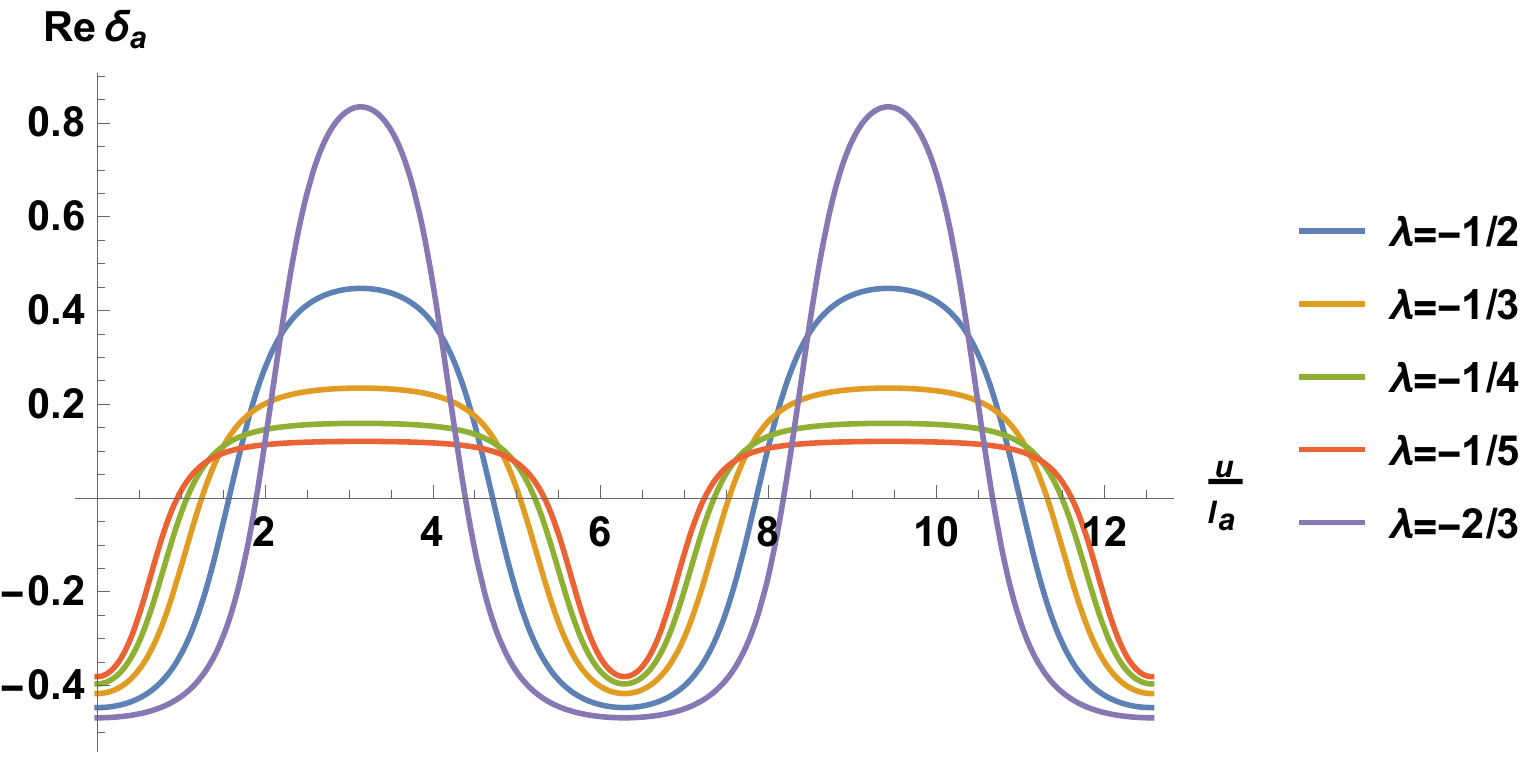}
\includegraphics[scale=0.4]{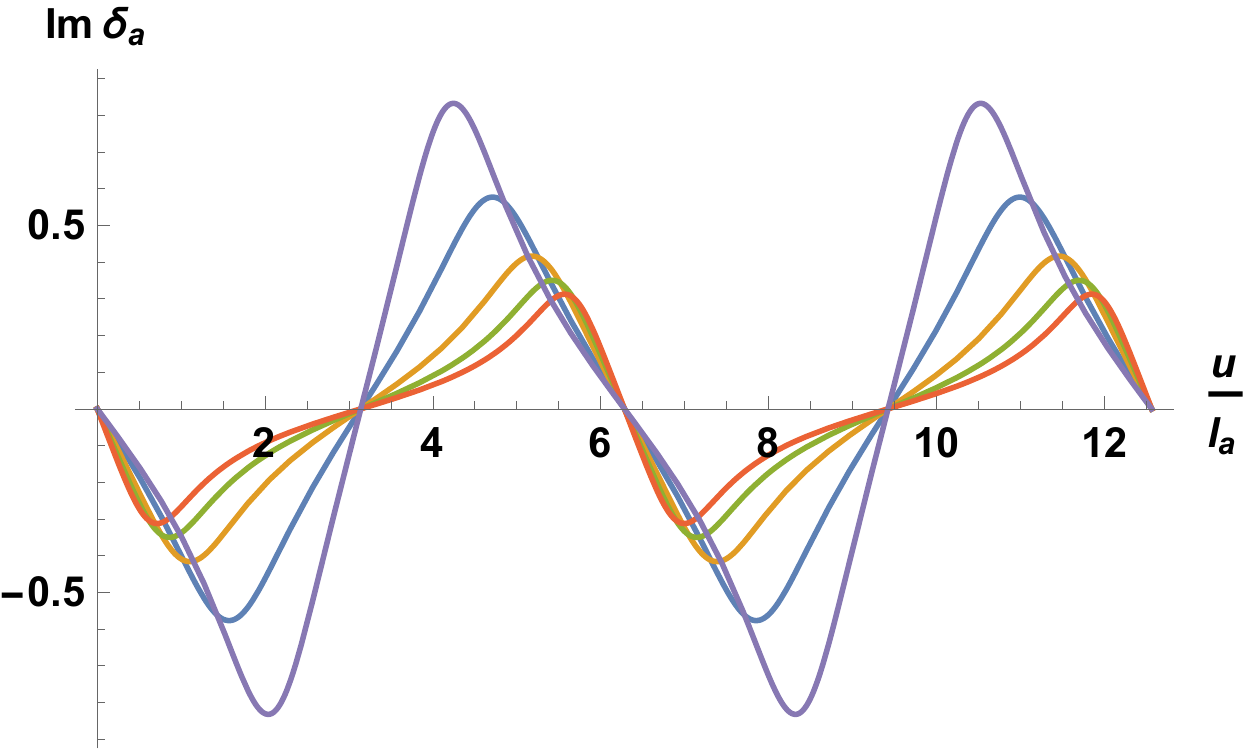}
\includegraphics[scale=0.4]{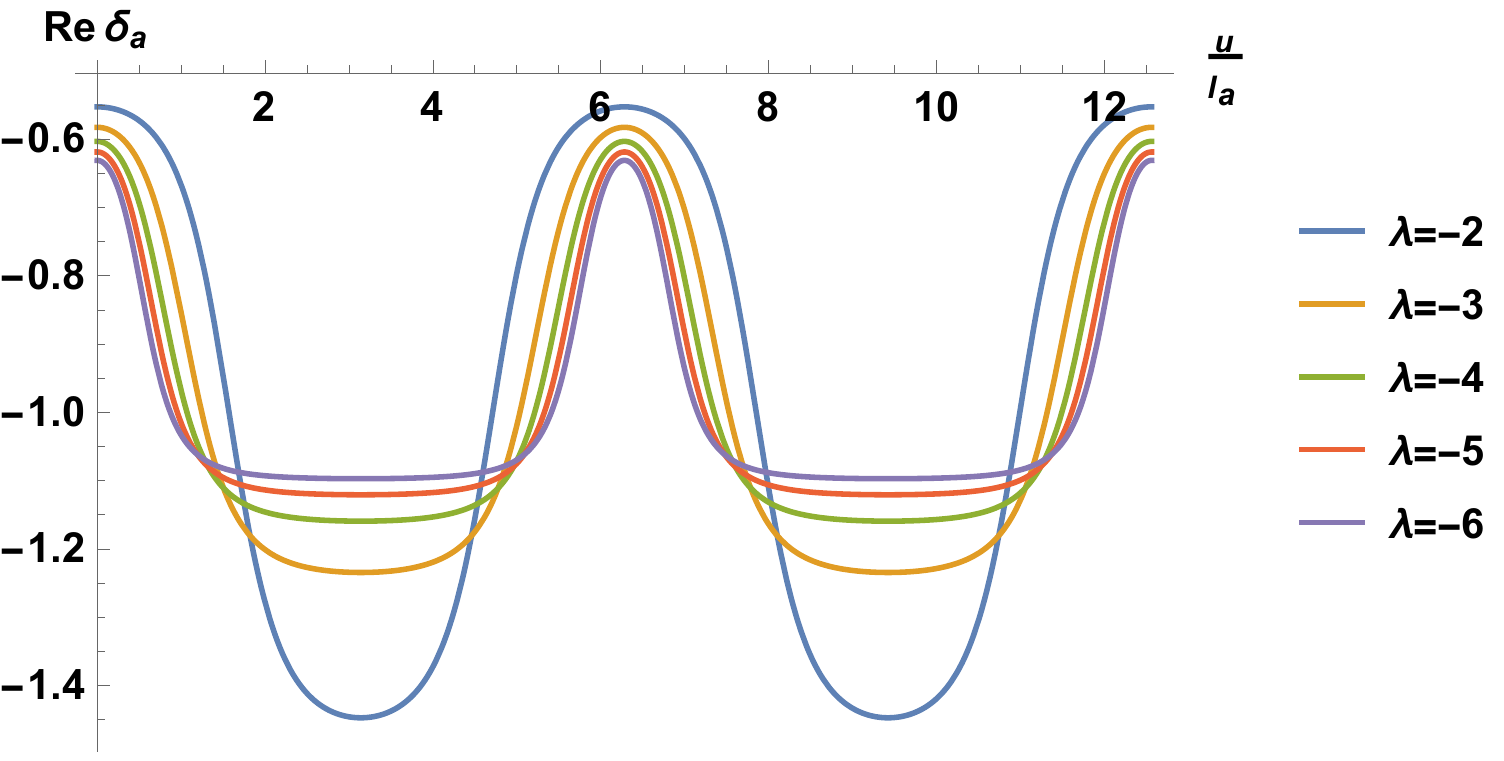}
\includegraphics[scale=0.4]{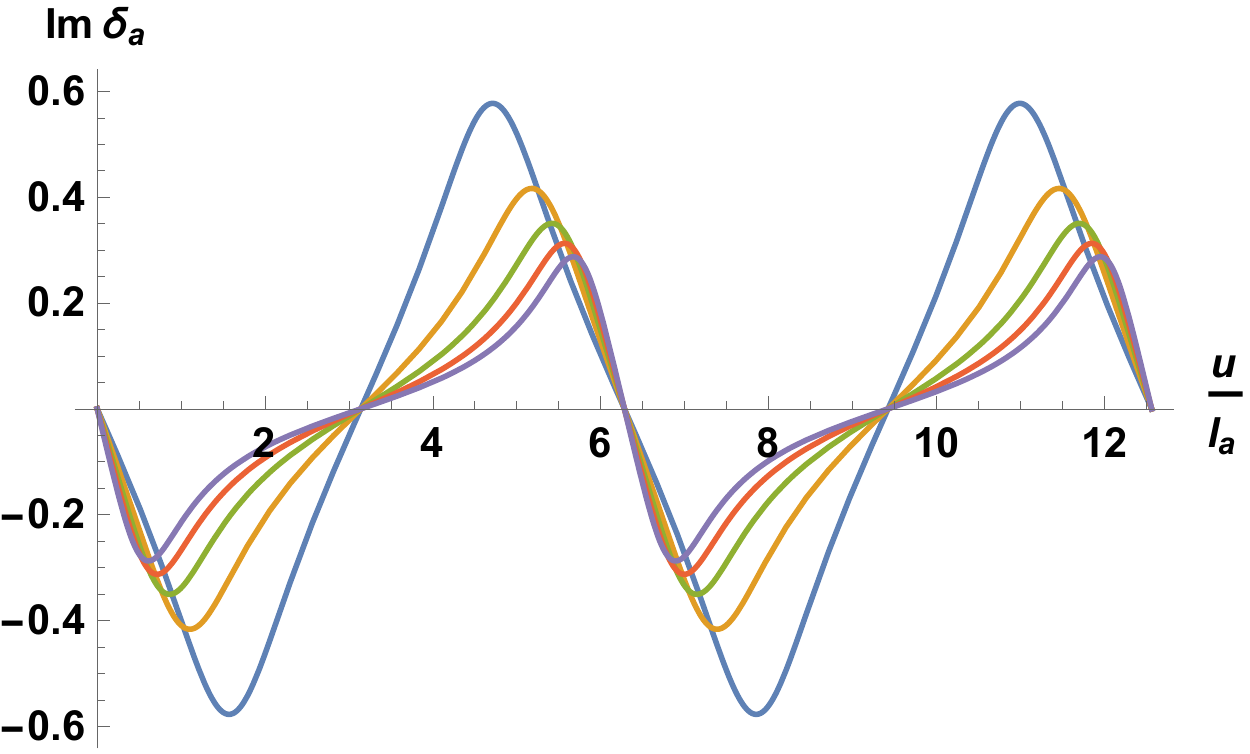}
\label{RatKinPlots}
\caption{Real and imaginary part of the string correction to the EM memory for some rational kinematical values.}
\end{figure}

In the CoM frame, one has $\vec{p}_1{=}\vec{p}{=}{-}\vec{p}_2$, $\vec{p}_3{=}{-}\vec{k}{=}{-}\omega\vec{n}$, so that $E_3{=} E^{\rm phys}_3{=}\sqrt{M_3^2{+}\omega^2}$, while  $E_{1,2}{=}{-}E^{\rm phys}_{1,2}{=}\sqrt{M_{1,2}^2{+}|\vec{p}|^2}$ with  
\be
E^{\rm phys}_1 = {\widetilde{M}_3^2+M_1^2-M_2^2 \over 2\widetilde{M}_3} \,, \qquad E^{\rm phys}_2 = {\widetilde{M}_3^2+M_2^2-M_1^2 \over 2\widetilde{M}_3} \,, \qquad
|\vec{p}| = {\sqrt{{\cal F}(M^2_1,M^2_2,\widetilde{M}^2_3)}\over 2\widetilde{M}_3} \,,
\ee
where $\widetilde{M}_3 = E_3{+}\omega$ and ${\cal F}(x,y,z)= x^2+y^2+z^2-2xy-2yz-2zx$, a.k.a. as the `fake square'.  Setting $\mu_1=M_1^2/\widetilde{M}_3^2$, $\mu_2=M_2^2/\widetilde{M}_3^2$ and $\cos\theta =  {\vec{x}{\cdot} \vec{p}\over{4\pi}R|\vec{p}|}$, ${\cal F}$ is positive in the physical domain:
$0{<}\mu_1,\mu_2{<}1$, $(\mu_1{-}\mu_2)^2{-}2(\mu_1{+}\mu_2){+}1{>}0$. 
In our example $\mu_1=\mu_2=\mu=M^2/\widetilde{M}_3^2$, one has 
\be
\lambda_{1,3} = -{1\over 2}+ {1\over 2} \cos\theta \sqrt{1- 4\mu} 
\, , \qquad \qquad  
\lambda_{2,3} = -{1\over 2} - {1\over 2} \cos\theta\sqrt{1- 4\mu} \,,
\ee
In particular, for $\cos\theta = 0$, 
$\ell_{1}=\ell_{2}=-\frac{1}{2}\ell_{3}$ so that
\be
\lambda_{1,3}\equiv\lambda_{2,3}=-\frac{1}{2}\,,\quad\lambda_{3,1}\equiv\lambda_{3,2}=-2\,,\quad\lambda_{1,2}\equiv\lambda_{2,1}=1\,.
\label{lambdaSpec}
\ee
Inserting these special values of $\lambda_{a,b}$ in \eqref{A>} and \eqref{A<}, for $u>0$ yields
\be
\begin{aligned}
\Delta^{(>)}_s A^{\mu}(t, \vec{x}) &= -{{g} \over{4\pi}R}\left({p^{\mu}_1\over n{\cdot}p_1} + {p^{\mu}_2\over n{\cdot}p_2} - 2{p^{\mu}_3\over n{\cdot}p_3} \right)\sum_{n_3=1}^\infty 
{(-)^{n_3} \over n_3!} {\Gamma(1{+}{n_3\over2})
\over  
\Gamma(1{-}{n_3\over 2})} \,e^{{{i}} n_3 u/{\ell}_3}\\
&={{g} \over{4\pi}R}\left({p^{\mu}_1\over n{\cdot}p_1} + {p^{\mu}_2\over n{\cdot}p_2} - 2{p^{\mu}_3\over n{\cdot}p_3} \right){e^{{{i}} u/{\ell}_3} \over \sqrt{4 + e^{2{{i}}u/{\ell}_3}}}
\end{aligned} 
\ee
with $n{{\cdot}}p_j = - E_j\left(1-{\vec{x}\vec{v}_j\over{4\pi}R} \right)$, and for $u<0$ one has 
\be
\begin{aligned}
\Delta^{(<)}_s A^{\mu}(t, \vec{x}) &={{g}\over{4\pi}R}\left({p^{\mu}_1\over n{\cdot}p_1} + {p^{\mu}_2\over n{\cdot}p_2} - 2{p^{\mu}_3\over n{\cdot}p_3} \right) \sum_{n_1=1}^\infty 
{(-)^{n_1} \over n_1!} {\Gamma(1{+}{2n_1})
\over \Gamma(1{+}n_1)} e^{{{i}} n_1 u/{{{\ell}}_1}}\\
&= {{g}\over{4\pi}R}\left({p^{\mu}_1\over n{\cdot}p_1} + {p^{\mu}_2\over n{\cdot}p_2} - 2{p^{\mu}_3\over n{\cdot}p_3} \right)\left({1 \over \sqrt{1+ 4e^{{{i}} u/{{{\ell}}_1}}}} - 1\right)
\end{aligned}
\ee
Although the radii of convergence of the series are finite ($|{{\zeta}}_3|<2$, $|{{\zeta}}_1|<1/4$) as for the closed (heterotic) string \cite{ABFMhet}, the explicit expressions allow to analytically continue both functions to the physical range $|{{\zeta}}_j|=1$ without the log terms found in \cite{ABFMhet}. In fact $\Delta^{(>)}_s{{A}}^\mu+\Delta^{(<)}_s{{A}}^\mu=0$ as a remnant of planar duality. For other choices of `rational' kinematics as the ones in Table~\ref{RatKinTab}, one finds similar results. We display the plots of the real and imaginary part of the string corrections to the EM wave profile at fixed large $R$ as a function of $u/\ell$ for some `rational' choices of the kinematics in Fig.~\ref{RatKinPlots}. As evident, in order to detect the string memory effect one needs a time resolution $\Delta{t}\approx \ell$.

\begin{table}[h!]
\centering
\begin{tabular}{cccccc}
 $\lambda_{13}$& $\lambda_{23}$ &$\lambda_{31}$&
 $\lambda_{21}$  & $\lambda_{12}$ &$\lambda_{32}$\\
\hline
-1/2 &-1/2& -2 &1& 1 &-2 \\
  -1/3 &-2/3& -3 &2& 1/2 &-3/2 \\
  -1/4 &-3/4& -4 &3& 1/3 &-4/3 \\
  -1/5 &-4/5& -5 &4& 1/4 &-5/4 \\
  -2/3 &-1/3& -3/2 &1/2& 2 &-3 \\
  -3/4 &-1/4& -4/3 &1/3& 3 &-4 \\
\end{tabular}
\caption{\label{RatKinTab} Some examples of `rational' kinematical regimes.}
\end{table}

In the very-high-energy limit $u/|\ell_j|<<1$, there are two possible regimes: fixed angle for $|\ell_1|\approx |\ell_2|\approx |\ell_3|$ and Regge $|\ell_1|<< |\ell_2|\approx |\ell_3|$. In the former case, as visible in the plots in Fig.~\ref{RatKinPlots}, the real part of the signal gets a constant shift, while the imaginary part has a linear behaviour in $u$. In the latter case, the real part has a long plateau in $u$ and the imaginary part has a sudden jump and then flattens down. These peculiar features of the stringy signal should allow to discriminate it from the standard EM memory or other field-theory effects. 


\section{\emph {Generalizations}}
One can generalise the analysis to higher-points and higher-loops or to more realistic models with open and unoriented strings.

\subsection{\emph{n-points}}
 Compatibly with $U(1)$ charge conservation, the non-vanishing scattering amplitudes involving photons and `tachyons' in our $U(1){{\times}}O(1)$ model are 
\be
{\cal A}[A_1, ..., A_{n_A}; \phi^+_1,...,\phi^+_{n_1}; \phi^-_1,...,\phi^-_{\bar{n}_1}; \chi^{+2}_1,...,\chi^{+2}_{n_2}; \chi^{{-}2}_1,...,\chi^{{-}2}_{\bar{n}_2}]
\ee
with $n_{1}{-}\bar{n}_{1}{+}2n_{2}{-}2\bar{n}_{2}{=}0$. Due to `twist' symmetry, $n_A$ must be even if $n_{1}{=}\bar{n}_{1}{=}n_{2}{=}\bar{n}_{2}{=}0$. 

\begin{figure}[h!]
\centering
\includegraphics[scale=0.3]{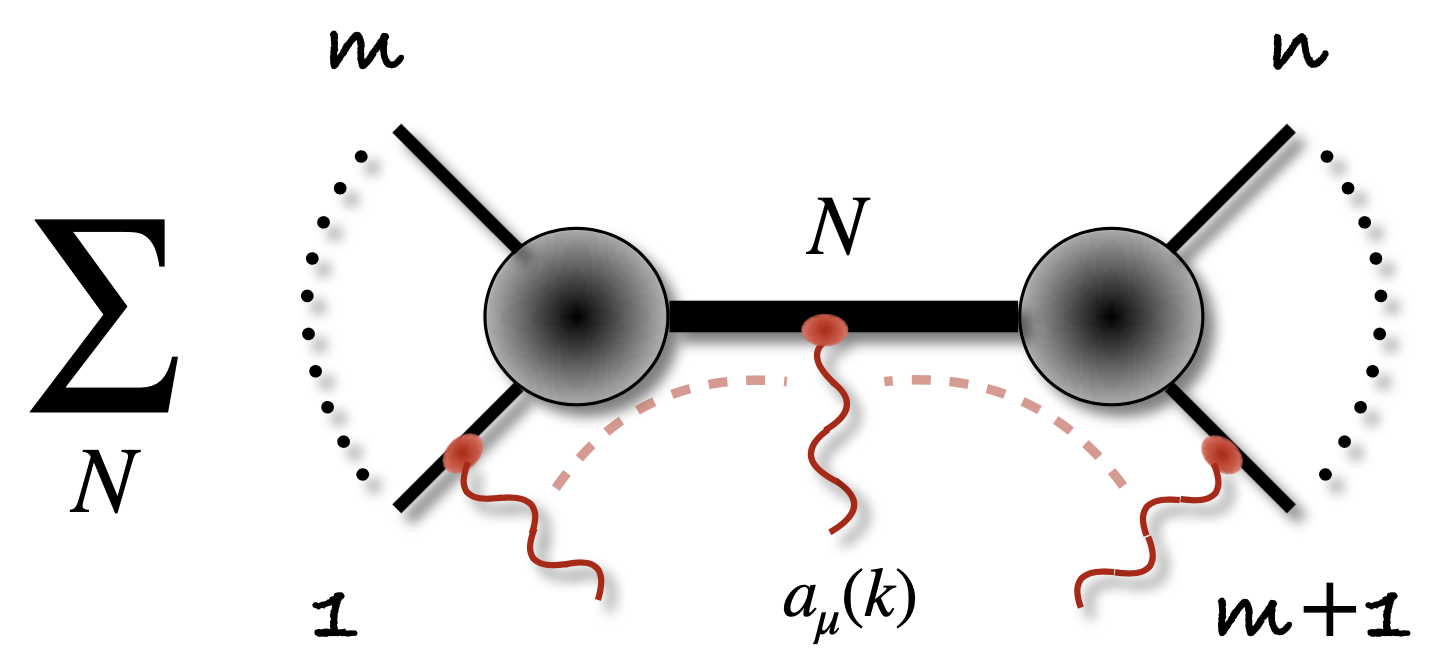}
\caption{Schematic picture of the factorized amplitude respect to the different classes of poles.}
\label{HigherPointFact}
\end{figure}

We are interested in the case $n_A=1$. The photon can be inserted on the $U(1)$ end of any of the $n$ `tachyon' legs.  For a given color-ordering at tree level, one has  
\be
{\cal A}^{c.o.}_{n{+}1}(a,k;p_j) = g_{op}^{n{-}1}(2\alpha')^{{n{-}2\over 2}}\int {\prod_{j=0}^n dz_j \over V_{CKV}} \sum_{i=1}^n {a{\cdot}p_i \over z_0{-}z_i} \prod_{j=1}^{n} z_{0j}^{2\alpha' kp_j}\prod_{i<j}^{1,n} z_{ij}^{2\alpha' p_ip_j}
\ee
with $z_{{{j}}}<z_0<z_{{{j}}{+}1}$ for some ${{{j}}}$. Integrating over $z_0$ near $z_{{{j}}}$ ($z_{{{j}}{+}1}$) produces  a soft pole $1/2\alpha' kp_{{{j}}}$ ($1/2\alpha' kp_{{{j}}+1}$) and a series of `massive' ones. Additional massive poles come from multi-particle channels involving the photon. These are present in QED and arise from photons inserted on internal lines. No soft poles are exposed for generic choices of the hard momenta since 
\be
{1\over (k+p_1+ .... p_m)^2 + M^2} = {1\over 2k(p_1+ .... p_m) +(p_1+ .... p_m)^2 + M^2}
\ee
has poles at $2\omega n P = - P^2 - M^2$ with $P=\sum_{i=1}^m p_i$, see Fig.~\ref{HigherPointFact}. While in the soft limit such terms are subleading in $\omega$
\be
{1\over 2k(p_1+ .... p_m) +(p_1+ .... p_m)^2 + M^2} \approx {1\over (p_1+ .... p_m)^2 + M^2} - {2\omega n(p_1+ .... p_m)\over [(p_1+ .... p_m)^2 + M^2]^2} + ....
\ee
integrating over $\omega$ a string amplitude produces series in ${{\zeta}}_P =\exp{i u\over 2\alpha' nP}$, arising from the infinite tower of intermediate states, which are the hallmark of the string memory and represent a novelty w.r.t. to the 4-point case.

\begin{figure}[h!]
\centering
\includegraphics[scale=0.4]{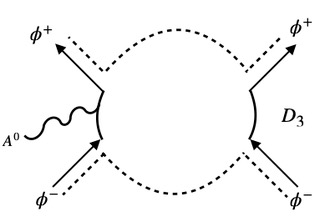}
\includegraphics[scale=0.4]{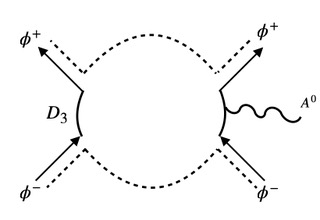}
\caption{Two contributions to the 5-point amplitude ${\cal A}_{4{+}1}[\phi^+(1), A(k), \phi^-(2), \phi^-(3), \phi^+(4)]$.}
\label{amp5pt}
\end{figure}

For instance, in our $U(1){{\times}}O(1)$ model, one can consider the 5-point amplitude in Fig.~\ref{amp5pt}
\be
{\cal A}_{4{+}1}[\phi^+(1), A(k), \phi^-(2), \phi^-(3), \phi^+(4)] 
\ee

the full amplitude includes insertions of $A(k)$ in between $\phi^+(3)$ and $\phi^-(4)$ as well as the exchanges $\phi^+(1)\leftrightarrow \phi^+(4)$ and 
$\phi^-(2)\leftrightarrow \phi^-(3)$. Let us focus on the indicated `color ordering' $[\phi^+(1), A(k), \phi^-(2), \phi^-(3), \phi^+(4)]$. 
Setting $z_4=\infty$, $z_3=0$ and 
$z_1=1$ with $0<z_A= z<1$ and $0<z_2= yz<z$ ({\it i.e.} $0<y<1$) one has
\be
{\cal A}^{\rm c.o.}_{4{+}1}[1^+, a(k), 2^-,3^+,4^-]=
2\alpha' g_{op}^3 \int_0^1 dz \int_0^1 dy \left( {a{{\cdot}}p_1\over 1-z}  - {a{{\cdot}}p_2\over z(1-y)} -{a{{\cdot}}p_3\over z}\right) {\times}
\ee
$$
(1-z)^{2\alpha'kp_1} (1-yz)^{2\alpha' p_1p_2} (1-y)^{2\alpha' kp_2} 
z^{2\alpha' (kp_2+kp_3+p_2p_3)+1} y^{2\alpha' p_2p_3} \, .
$$
Expanding $(1-yz)^{2\alpha' p_1p_2} = \sum_{N=0}^\infty (-2\alpha' p_1p_2)_Ny^N z^N/N!$, 
the integrals in $z$  and $y$ decouple. Setting $2\alpha'=1$ and using momentum conservation one can factor out the function
\be 
{\cal F}_N(k,p_j) = {\Gamma(p_2p_3 {+} k(p_{2}{+}p_{3}){+}N{+}1)\Gamma(kp_1{+}1)
\over \Gamma(p_1p_4{-}kp_{3}{+}N{+}1)} {\Gamma(p_1p_4{+}k(p_{1}{+}p_{4}){+}N{+}1)\Gamma(kp_2{+}1)
\over \Gamma(p_2p_3{-}kp_{4}{+}N{+}1)}
\ee
and get  
\be
\begin{aligned}
{\cal A}^{\rm c.o.}_{4{+}1}&[1^+, a(k), 2^-,3^+,4^-]=
 g_{op}^3 \sum_{N=0}^\infty {(- p_3p_4 - k(p_{3}{+}p_{4}))_N \over N!} {\cal F}_N(k,p_a) {\times}\\
&\bigg[ {a{{\cdot}}p_1(p_{2}p_{3}{+}k(p_{3}{+}p_{3}){+}N{+}1) \over kp_1(p_{1}p_{4}{-}kp_3{+}N{+}1)(p_{2}p_{3}{-}kp_4{+}N{+}1)}\\
&-{a{{\cdot}}p_2 \over kp_2  (p_2p_3{-}kp_{4}{+}N{+1})} -
{a{{\cdot}}p_3  \over (p_2p_3{-}kp_{4}{+}N{+1})(p_1p_4{-}kp_{3}{+}N{+1})}\bigg]\,.
\end{aligned}
\ee
Thanks to ${\cal F}_N(k,p_a)\big|_{\omega=0}=1$, the leading behavior at the soft pole ${\omega=0}$ is as expected. 
In addition there are four infinite sets of simple poles at
\be
\omega_{n_1} = - {n_1{+}1\over {{\ell}}_1} 
\quad 
\omega_{n_2} = - {n_2{+}1\over {{\ell}}_2} 
\quad 
\omega_{n_{23}} = - {n_{23}{+}1{+}N{+}p_2p_3\over {{\ell}}_2{+}{{\ell}}_3} \quad 
\omega_{n_{14}} = - {n_{14}{+}1{+}N{+}p_1p_4\over {{\ell}}_1{+}{{\ell}}_4}
\ee
The first two correspond to photon insertion on an external `tachyon' leg, as in the case of 4-point amplitudes, the last two to insertion on an `internal' leg (multi-particle channel)\footnote{Note that the Pochhammer symbol $(-p_ip_j)_N= \Gamma(- p_ip_j{+}N)/\Gamma(- p_ip_j)$ is a polynomial of degree $N$ and has no pole in its argument.}, a novel feature of 5- and higher-point amplitudes.
Using the ML expansion, the string corrections to $A_\mu(t, \vec{x})$ at large distance  read
\be
\Delta_s{A}^{\mu}(t, \vec{x})= {g^{3}_{op}\over{4\pi}R} \sum_{N=0}^\infty \left[
\widehat{\cal S}^{(1)}_N({{\zeta}}_1) + \widehat{\cal S}^{(2)}_N({{\zeta}}_{2}) + \widehat{\cal S}^{(23)}_N({{\zeta}}_{23})  + \widehat{\cal S}^{(14)}_N ({{\zeta}}_{14})\right]
\ee
where ${{\zeta}}_1 = e^{+ {{i}} {u\over {{\ell}}_1}} $, ${{\zeta}}_2 = e^{+ {{i}} {u\over {{\ell}}_2}}$, ${{\zeta}}_{23} = e^{+ {{i}} {u\over {{\ell}}_2 +{{\ell}}_3}} $, ${{\zeta}}_{14} = e^{+ {{i}} {u\over {{\ell}}_1 +{{\ell}}_4}}$
and 
\be
\begin{aligned}
\widehat{\cal S}^{(1)}_N({{\zeta}}_1) = & \sum_{n_1=1}^\infty{(-)^{n_1} {{\zeta}}_1^{n_1}\over n_1!}{\left(-p_{3}p_{4}{+}{n_{1}(\ell_{3}{+}\ell_{4})\over \ell_{1}}\right)_{N}\over N!}\Bigg[{a{\cdot}p_{1}\left(p_{2}p_{3}{-}{n_{1}\left(\ell_{2}{+}\ell_{3}\right)\over \ell_{1}}{+}N{+}1\right)\over\left(p_{2}p_{3}+n_{1}\frac{\ell_{4}}{\ell_{1}}{+}N{+}1\right)\left(p_{1}p_{4}+n_{1}\frac{\ell_{3}}{\ell_{1}}{+}N{+}1\right)}\\
&-{a{\cdot}p_{2}\over\left(p_{2}p_{3}+n_{1}\frac{\ell_{4}}{\ell_{1}}{+}N{+}1\right){\ell_{2}\over\ell_{1}}}{+}{a{\cdot}p_{3}\, n_{1}\over \left(p_{2}p_{3}+n_{1}\frac{\ell_{4}}{\ell_{1}}{+}N{+}1\right)\left(p_{1}p_{4}+n_{1}\frac{\ell_{3}}{\ell_{1}}{+}N{+}1\right)}
\Bigg]\\
&{\Gamma\left(p_{1}p_{4}-n_{1}\left(1{+}\frac{\ell_{4}}{\ell_{1}}\right){+}N{+}1\right)
\over \Gamma\left(p_{1}p_{4}+n_{1}\frac{\ell_{3}}{\ell_{1}}{+}N{+}1\right)} {\Gamma\left(p_{2}p_{3}-n_{1}\frac{(\ell_{2}{+}\ell_{3})}{\ell_{1}}{+}N{+}1\right)\Gamma(1{-}n_1{{{\ell}}_2\over{{\ell}}_1})
\over  \Gamma\left(p_{2}p_{3}+n_{1}\frac{\ell_{4}}{\ell_{1}}{+}N{+}1\right)}
\end{aligned}
\ee
Similar expressions obtain for $\widehat{\cal S}^{(2)}_N({{\zeta}}_{2})$, $\widehat{\cal S}^{(23)}_N({{\zeta}}_{23})$ and $\widehat{\cal S}^{(14)}_N ({{\zeta}}_{14})$. In addition one has to include the contributions arising from different `color orderings'.
The double series $\widehat{\cal S}^{(i)}_N({{\zeta}}_{i})$ look hard to write in closed form. Assuming particle 1 and 4 to be incoming and 2 and 3 as well as the photon to be outgoing, the simplest choice of kinematics is taking $\ell_1=\ell_4=-\ell_2=-\ell_3$.
In the CoM frame, with $p_1=-(E,\vec{p})$, $p_4=-(E,-\vec{p})$, $k=\omega(1,\vec{n})$, $p_2=-(E_2,\vec{p}_2)$, $p_4=-(E_3,\vec{p}_3)$, 
one has $2E=   \omega+E_2+E_3$, $\vec{p}_2+\vec{p}_3=-\omega\vec{n}$. $\ell_1=\ell_4$ implies photon emission perpendicular to the incident beams ($\vec{n}{{\cdot}}\vec{p}=0$). Then $\ell_2=\ell_3$ implies $E_2=E_3=E-{\omega \over 2}$ and $\vec{p}_{2,3}{=}{\pm}\vec{q}{-}{\omega \over 2}\vec{n}$, with $\vec{n}{{\cdot}}\vec{q}=0$. 
In this special kinematic regime, setting 
$\ell{=}2\alpha'E$ one finds 
${{\zeta}}_1{=}{{\zeta}}_2^{-1}{=}{{\zeta}}_{23}^{-2}{=}{{\zeta}}_{14}^{2}{=}\exp(iu/\ell)$.  Moreover $p_1p_2=p_3p_4=\hat{t}$, $p_1p_3=p_2p_4=\hat{u}$ and $p_1p_4+\omega\ell= \hat{s}=p_2p_3-\omega\ell$ with $\omega\ell=-n_i$ and $\hat{s}{+}\hat{t}{+}\hat{u}=0$. Even in this simple case, unless one further specialises the remaining 4-point kinematics, the final expressions are not very illuminating and we refrain from displaying them here\footnote{They will appear in the companion paper \cite{Compa}.}. Moreover the ratios of the functions ${\cal S}(\zeta)$ to the residual 4-point `tachyon' amplitude depend on
$\hat{s}$, $\hat{t}$ and $\hat{u}$.

\subsection{\emph{Loops}}
One-loop annulus and M\"obius-strip amplitudes can be written down and analysed in general and in particular in the $U(1){{\times}}O(1)$ model. The combinatorics of boundaries is more involved than at the disk level. So let us focus on the planar case, whereby all the insertions are on the same boundary of an annulus. For a given color-ordering, one has  
\be
{\cal A}^{c.o.}_{n{+}1}(a,k;p_j) = {\cal C}_A g_{op}^{n{+}1}(2\alpha')^{{n{+}2\over 2}}\int_0^\infty {d\tau\over\tau^{14}\eta(\tau)^{24}} \int {\prod_{j=0}^n dz_j \over V_{CKV}} 
\sum_{j=1}^n a{\cdot}p_j \partial_{z_0} {\cal G}(z_0{-}z_j){\times}
\ee
$$ 
\prod_{j=1}^{n} \exp\{-2\alpha' kp_j{\cal G}(z_0{-}z_j)\}
\prod_{i<j}^{1,n} \exp\{-2\alpha' p_ip_j{\cal G}(z_i{-}z_j)\}
$$
where $\eta(\tau)$ is Dedekind function, ${\cal G}(z) = -\log[\vartheta_1(z)/\vartheta'_1(0)] - (\pi z^2/\tau)$ is the bosonic propagator on the annulus with $\tau{=}i\tau_2$. The moduli space includes loci where the annulus degenerates. The UV region $\tau{\rightarrow} 0$ produces the closed-string tachyon pole (off-shell) and the dilaton tadpole that cancels for $SO(8192)$ or its Wilson-line breakings. When $z_0$ collides with the adjacent points one gets log-`deformed' poles giving rise to the one-loop corrections to the EM memory and to the string memories. When all but one of the points collide one gets the one-loop corrections to the mass and width of the particle. Other degenerations that start to appear at higher loops produce quantum corrections to lower-point amplitudes that require a case-by-case analysis but do not affect our main result in so fa as $g_s<<1$. 

\subsection{\emph{Type I models}}
In more realistic models with open and unoriented superstrings, that allow an embedding of the Standard Model or (supersymmetric) extensions thereof, $U(1)$ gauge bosons are typically anomalous and massive\footnote{For recent work on axions, (gravi-)photons and their mixings in holographic setups see e.g. \cite{AxionGravPhot}.}  \cite{U1anommass}. Yet, massless non-anomalous combinations exist that can play the role of the SM hyper-charge \cite{YinString}. {\it Mutatis mutandis} and barring issues such as moduli stabilisation and supersymmetry breaking, open superstring amplitudes with photons and charged scalars or fermions present the same structure as in our $O(1){\times}U(1)$ bosonic model and the analysis of the EM String Memory proceeds along the same steps. 

Different (holographic) scenari are possible \cite{HoloQCD}, whereby QCD is described by some strongly coupled sector (`color branes') and the electro-weak sector is coded in a weakly coupled sector (`flavour branes'). These configurations involve large fluxes and strong warping that lower the effective string tension. Although we expect genuine string corrections to EM memory also in these contexts, at present we cannot support our statement with explicit computations of the relevant amplitudes.

{\bf \emph {Discussion.}} While the coherent effect of the infinite tower of string resonances may well give rise to a detectable modulated EM `memory' we have not given any formal argument why it should happen altogether. Our results do not point towards a DC effect as in the standard memory in gravity and in EM \cite{OldEMMemo, OldGravMemo, NewMemo}. Neither they seem to be related to the proposed 'string memory effect' that involve large gauge transformations of the Kalb-Ramond field \cite{Afshar1811.07368}. Instead the effect we find is oscillatory and it is tempting to conjecture that it be related to the `global' part of the infinite (but broken) higher spin symmetries of string theory. In particular, redundancies in the definition of the physical states,  $\Psi \sim \Psi+{\cal Q}_{BRST} \Lambda$ involved in the scattering process may effect the `gauge' chosen for the incoming string states w.r.t. the gauge chosen for the outgoing string states. Note that not only states in the first Regge trajectory but also states with mixed symmetry and lower spin admit complicated `gauge symmetries' that may be exposed in certain regimes \cite{HigherSpinNoi, HSalter, HighEnergy}.

Moreover, in a given model, only for a special range of the parameters and the masses of the particles involved in the collision the effect may give rise to a measurable signal. 
A reasonable time-scale, compatible with present detector resolutions, would be $\Delta{t} \approx \alpha' E \approx 10^{-15}s = 1 fs$  that in turn would require $E\approx 10^{15}GeV$, well beyond the present and near future accelerators even for $TeV$-scale (super)strings with 
$\alpha'=E_s^{-2}\approx 10^{-6} GeV^{-2}$ \cite{TeVStrings}. An alternative more promising scenario is a collision of macroscopic objects such as (open) string coherent states \cite{MBMF, AAMF, Skliros} with large mass (and spin) $E>>10^{15}GeV$. In this case, even for $\Delta{t} >> 1 fs$ one could hope to recognise the string imprints in the EM radiation.

\section*{Acknowledgements}

We would like to thank A.~Addazi, A.~Anastasopoulos, D.~Consoli,  F.~Fucito,  A.~Grillo,  Y-t~Huang, E.~Kiritsis,  S.~Mancani, A.~Marcian\'o,  J.~F.~Morales, G.~Pradisi, G.~C.~Rossi, R.~Russo, R.~Savelli, C.~Sonnenschein, N.~Tantalo,  P.~Di~Vecchia, G.~Veneziano for interesting discussions and valuable comments on the manuscript. 
The work is partly supported by the University of Tor Vergata through the Grant ``Strong Interactions: from Lattice QCD to Strings, Branes and Holography'' within the Excellence Scheme ``Beyond the Borders''.

\end{document}